\pdfoutput=1
\documentclass[final]{IEEEtran}

\usepackage{amsmath,amsfonts,graphicx,amssymb,cite,bm,multirow,multicol,subfigure}

\usepackage{algorithm}
\usepackage{algorithmic}

\usepackage{url}
\makeatletter
\def\url@leostyle{%
  \@ifundefined{selectfont}{\def\UrlFont{\sf}}{\def\UrlFont{\small\ttfamily}}}
\makeatother
\newtheorem{lemma}{Lemma}
\newtheorem{theorem}{Theorem}
\newtheorem{rem}{Remark}

\newcommand{\vectorsymbol}{\underline}
\newcommand{\matrixsymbol}{\boldsymbol}

\newcommand{\mA}{\matrixsymbol{A}}
\newcommand{\mC}{\matrixsymbol{C}}
\newcommand{\mhC}{\widehat{\matrixsymbol{C}}}
\newcommand{\mCs}{\matrixsymbol{C}_\text{S}}
\newcommand{\mI}{\matrixsymbol{I}}

\newcommand{\mW}{\matrixsymbol{W}}

\newcommand{\mhW}{\widehat{\matrixsymbol{W}}}
\newcommand{\mL}{\matrixsymbol{L}}
\newcommand{\mLambda}{\matrixsymbol{\Lambda}}

\newcommand{\va}{\vectorsymbol{a}}
\newcommand{\vn}{\vectorsymbol{n}}
\newcommand{\vs}{\vectorsymbol{s}}
\newcommand{\vw}{\vectorsymbol{w}}
\newcommand{\vhw}{\hat{\vectorsymbol{w}}}
\newcommand{\vx}{\vectorsymbol{x}}

\newcommand{\ulambda}{\lambda_0}

\newcommand{\ce}{c_\text{E}}
\newcommand{\Ce}{C_\text{E}}
\newcommand{\ci}{c_\text{I}}

\newcommand{\reals}{\mathbb{R}}
\newcommand{\complexes}{\mathbb{C}}

\DeclareMathOperator*{\diag}{diag}

\newcommand{\E}{\mathbb{E}}
\newcommand{\Prob}{\mathbb{P}}

\newcommand{\half}{\frac{1}{2}}

\DeclareMathOperator*{\Ai}{Ai}

\title{Minimax Rank Estimation for Subspace Tracking}
\author{Patrick~O.~Perry and~Patrick~J.~Wolfe,~\IEEEmembership{Senior Member,~IEEE}%
                \thanks{This material is based upon work
          supported in part by the Defense Advanced Research Projects Agency under Grant~HR0011-07-1-0007, the Army Research Office under Grant~W911NF-09-1-0555, and by the National Science Foundation under Grants~DMS-0604939 and 0652743.
          \newline \indent The authors are with the Statistics and Information Sciences Laboratory, Harvard University, Cambridge, MA 02138 USA (e-mail: \{patperry, patrick\}@seas.harvard.edu).
          }
}
\begin{document}

\maketitle

\begin{abstract}%
Rank estimation is a classical model order selection problem that arises in a variety of important statistical signal and array processing systems, yet is addressed relatively infrequently in the extant literature.  Here we present sample covariance asymptotics stemming from random matrix theory, and bring them to bear on the problem of optimal rank estimation in the context of the standard array observation model with additive white Gaussian noise.  The most significant of these results demonstrates the existence of a phase transition threshold, below which eigenvalues and associated eigenvectors of the sample covariance fail to provide any information on population eigenvalues.  We then develop a decision-theoretic rank estimation framework that leads to a simple ordered selection rule based on thresholding; in contrast to competing approaches, however, it admits asymptotic minimax optimality and is free of tuning parameters.  We analyze the asymptotic performance of our rank selection procedure and conclude with a brief simulation study demonstrating its practical efficacy in the context of subspace tracking.
\end{abstract}


\ifCLASSOPTIONpeerreview
\begin{center}%
\bfseries EDICS Category: SAM-PERF / SSP-PERF
\end{center}
\fi
\IEEEpeerreviewmaketitle

\section{Introduction}
\label{sec:intro}

\IEEEPARstart{R}{ank} estimation is a important model order selection problem that arises in a variety of critical engineering applications, most notably those associated with statistical signal and array processing.  Adaptive beamforming and subspace tracking provide two canonical examples in which one typically assumes $n$-dimensional observations that linearly decompose into a ``signal'' subspace of dimension $r \ll n$ and complementary ``noise'' subspace of dimension $n - r$.  In many applications the goal is to enhance, null, or track certain elements of the signal subspace, based on observed array data.

In this context, we of course recover the classical statistical trade-offs between goodness of fit and model order; i.e., between system performance and complexity.  In the rank selection case these trade-offs are particularly clear and compelling, as subspace rank estimation may well correspond to correctly identifying the number of interferers or signals of interest.  We note that the goal of this article is a general understanding of the model order selection problem in this context, rather than an exhaustive application-specific solution.

To this end, we present here a collection of recent results from random matrix theory, and show how they enable a theoretical analysis of this instantiation of the model order selection problem.  In Section~\ref{S:prob-state} we formulate the problem of model order selection  in the context of the standard array observation model with additive white Gaussian noise, and in Section~\ref{S:null-alt-rmt} we present sample covariance asymptotics based on random matrix theory.  In Section~\ref{S:rank-est} we bring these results to bear on the problem of optimal rank estimation by developing a decision-theoretic rank estimation framework, and associated algorithm whose asymptotic minimax optimality we prove.  We then provide a brief simulation study in Section~\ref{S:emp-perf} to demonstrate the practical efficacy of our rank selection procedure, and conclude with a summary discussion in Section~\ref{s:summ}.

\section{Problem Formulation: Model Order Selection}
\label{S:prob-state}

Suppose at time $t$ we observe data vector $\vx(t) \in \complexes^n$
(or $\reals^n$), a weighted combination of $r$ signal vectors corrupted
by additive noise.  Using the notation $\mathcal{M}_r$ to denote an additive observation model of order $r$, we can express this as
\begin{equation}\label{E:Signal}
    \mathcal{M}_r\!:\, \vx(t) = \sum_{i=1}^{r} s_{i}(t) \va_{i} + \vn(t) = \mA \vs(t) + \vn(t),
\end{equation}
where
\(
    \mA
    =
    \left(
    \begin{matrix}
        \va_{1} & \va_{2} & \cdots & \va_{r}
    \end{matrix}
    \right)
\)
is an $n \times r$ mixing matrix and
\(
    \vs(t)
    =
    \big(
        s_1(t), s_2(t), \ldots, s_r(t)
    \big)
\)
an $r \times 1$ signal vector.  Here, we assume that $\mA$ is a deterministic matrix of weights and
$\vs(t)$ is a random source vector with nonsingular covariance matrix $\mCs$.
If the noise $\vn(t)$ is independent of the source and white, with
all components having variance $\sigma^2$, then the covariance of $\vx(t)$ is
given by
\begin{equation}\label{E:C-decomp}
    \mC
    =
    \E\left[ \vx(t) \vx(t)^* \right]
    =
    \mA \mCs \mA^* + \sigma^2 \mI.
\end{equation}

In most array processing applications, it is generally desired to estimate or determine $\mA$ based on ``snapshot'' data vectors $\vx(t)$; however, note that the decomposition in~\eqref{E:C-decomp} is not statistically identifiable.  We can render it identifiable up to a sign change by reparametrizing as
\begin{equation}\label{E:C-decomp-eig}
    \mC
    =
    \mW \mLambda \mW^* + \sigma^2 \mI,
\end{equation}
with
\(
    \mW
    =
    \left(
    \begin{matrix}
        \vw_1 & \vw_2 & \cdots & \vw_r
    \end{matrix}
    \right)
\)
an $n \times r$ matrix having orthonormal columns and
\(
    \mLambda
    =
    \diag(
        \lambda_1, \lambda_2, \ldots, \lambda_r
    )
\)
a diagonal matrix with
\(
    \lambda_1 \geq \lambda_2 \geq \dots \geq \lambda_r > 0.
\)
Then, in many applications we can recover the parameters of interest from
$\mW$ using specialized knowledge of the structure of $\mA$ in conjunction with well-known algorithmic approaches
such as MUSIC~\cite{schmidt1986mel} or ESPRIT~\cite{roy1989ees}.

In general, then, the goal of subspace tracking is to estimate $\mW$ as well as
possible from observed array data $\vx$.  Often, the associated data covariance $\mC$ will change in time; signals may change their physical characteristics or direction of arrival, with some ceasing while others appear.  Such situations require \emph{adaptive} estimation of $\mW$.  Over the past twenty years, a variety of algorithms
have been developed for recursively updating estimates of the dominant subspace
of a sample covariance matrix.  Projection Approximation Subspace Tracking (PAST)~\cite{yang1995pas} is among the most popular, though many new algorithms continue to be proposed (see, e.g.,~\cite{badeau2008fas, bartelmaos2008fpc, doukopoulos2008fas}).

A deficiency of nearly all of these algorithms is that they require prior
knowledge of $r$, the rank of the desired signal subspace.  In relative terms, little
attention has been devoted in the extant literature to estimating $r$ in an optimal manner.  Kav\v{c}i\'c and Yang~\cite{kavcic1996are}, Rabideau~\cite{rabideau1996fra}, and Real et al.~\cite{real1999taf} separately address the problem, but ultimately all
rely on selection rules with problem-specific tuning parameters.  More recently,
Shi et al.~\cite{shi2007aen} suggest a modification of~\cite{rabideau1996fra}; each of these approaches seeks to derive appropriate tuning parameters based on
user-specified false alarm rates.

In parallel to the above developments, the literature on random matrix theory
has progressed substantially over the past ten years. An important outgrowth of this theory is
a new set of rank-selection tools.  To this end, Owen and Perry~\cite{owen2009bcv} suggest a
cross-validation-based approach that, for computational reasons, does not appear immediately applicable to real-time subspace tracking. Kritchman and
Nadler~\cite{kritchman2008dnc} provide a survey of other approaches, including those based on information-theoretic criteria~\cite{wax1985det, rao2008sam}, ultimately recommending a method based on eigenvalue thresholding. 
Their threshold is determined from a specified false alarm rate using the
theory developed in~\cite{johansson2000sfa, johnstone2001dle, baik2005pto, baik2006els, paul2007ase, onatski2007adp}.

Even for the rank selection procedures with interpretable tuning parameters such as false alarm rate, it is not clear how these parameters should be chosen.  In the sequel, we summarize recent sample covariance asymptotics and employ them to develop a decision-theoretic framework for estimation with specified costs for over- or under-estimating the rank.  Within this framework we derive a selection rule based on eigenvalue thresholding, without the need for tuning parameters, that minimizes the maximum risk under a set of suitable alternate models.

\section{Sample Covariance Asymptotics}
\label{S:null-alt-rmt}

Recall that the $r$th-order signal-plus-noise observation model $\mathcal{M}_r\!:\, \vx(t) = \mA \vs(t) + \vn(t)$ of~\eqref{E:Signal} gives rise to the covariance form $\mC$ defined in~\eqref{E:C-decomp} and~\eqref{E:C-decomp-eig}.  The rank selection rule we derive is based on a sample covariance matrix $\mhC$ comprised of $N$ array snapshots $\vx(t)$, indexed as a function of time $t$.  For $N \leq t$, this empirical covariance estimator is defined as
\begin{equation}\label{E:C-empirical}
    \mhC = \frac{1}{N} \sum_{k=0}^{N-1} \vx(t-k) \vx(t-k)^*,
\end{equation}
and our appeal to random matrix theory will rely on properties of $\mhC$ as the number of snapshots $N$ becomes large.

As is customary in the case of random matrix theory, we work in an asymptotic setting with $N$ and $n$ both tending to infinity, and their ratio $n/N$ tending to a constant $\gamma \in (0,\infty)$.  Moreover, we suppose that the number of signals $r$ is fixed with respect to this asymptotic setting, although it is likely that this assumption can in fact be relaxed to $r = o(\sqrt{n})$.  To simplify the presentation of the theory, we also suppose strict inequality in the ordering
$\lambda_1 > \lambda_2 > \cdots > \lambda_r > 0$ with respect to the decomposition of~\eqref{E:C-decomp-eig} of the true covariance $\mC$.  Note that under the assumed model of~\eqref{E:Signal}, the actual eigenvalues of $\mC$ are given by $\{\lambda_i+\sigma^2\}_{i=1}^r \,\cup\, \{\sigma^2\}_{i=r+1}^n$.

\begin{figure}[t]
    \centering
    \includegraphics[width=\columnwidth]{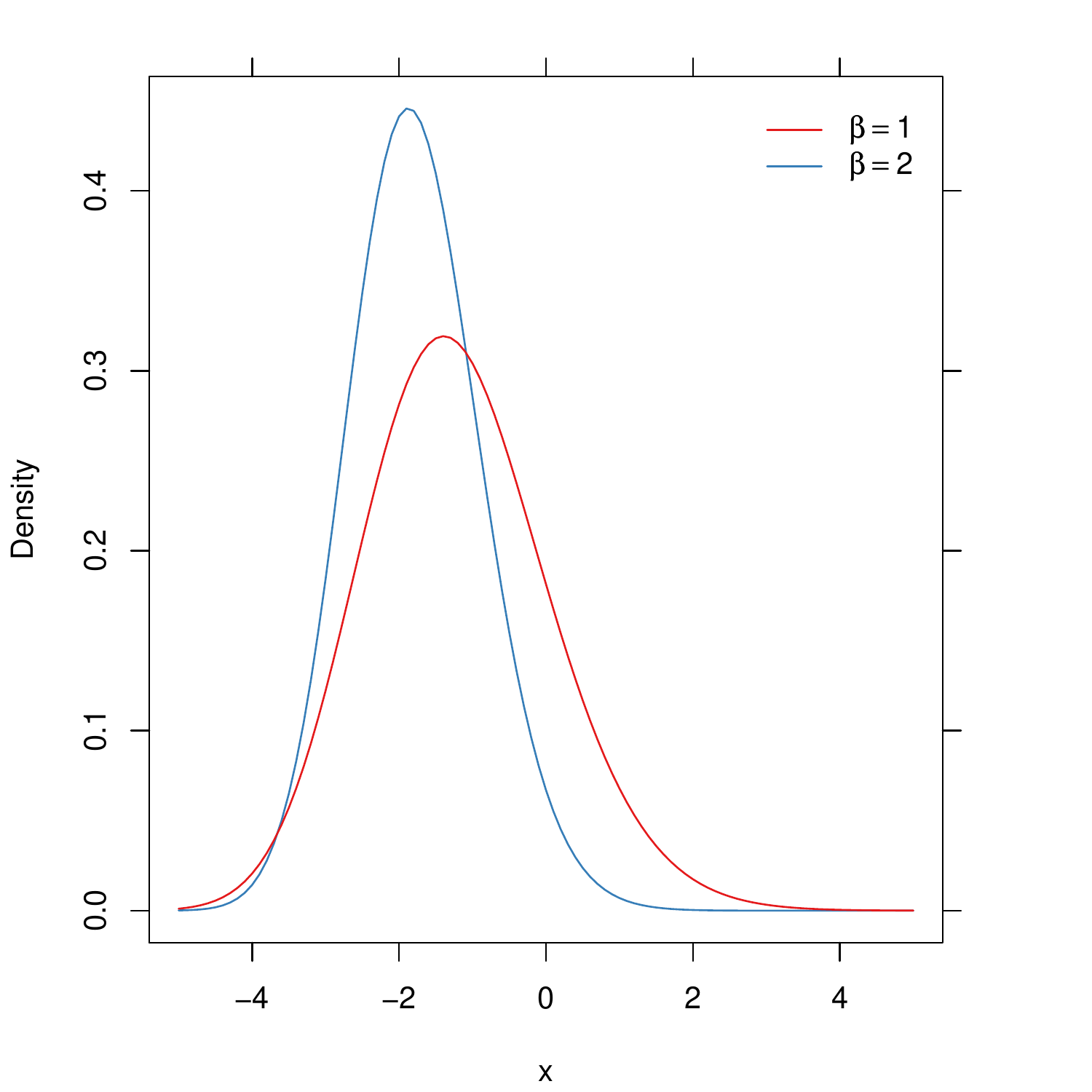}
    \caption{\label{F:tw-density}Density of the Tracy-Widom law $F_\beta(x)$ for real ($\beta=1$) and complex ($\beta=2$) cases, computed using the online software packages~\cite{dieng2006rml, perry2009rmtstat}.
    }
\end{figure}

To begin, we define the eigendecomposition of the empirical covariance $\mhC$ of~\eqref{E:C-empirical} as
\begin{equation}\label{E:Cemp-decomp-eig}
    \mhC = \mhW \mL \mhW^* \text{,}
\end{equation}
where
\(
    \mhW
    =
    \left(
    \begin{matrix}
        \vhw_1 & \vhw_2 & \cdots & \vhw_n
    \end{matrix}
    \right)
\)
has orthonormal columns and
\(
    \mL
    =
    \diag( \ell_1, \ell_2, \ldots, \ell_n )
\)
has $\ell_1 \geq \ell_2 \geq \cdots \geq \ell_{n} > 0$.  Now consider the additive observation model $\mathcal{M}_0$ of~\eqref{E:Signal} corresponding to $r=0$; i.e., in the \emph{absence} of signal:
\begin{equation*}
    \mathcal{M}_0\!:\, \vx(t) = \sum_{i=1}^{r} s_{i}(t) \va_{i} + \vn(t), \quad r = 0 .
\end{equation*}
This $0$th-order case defines a natural null model for our rank estimation problem, in that $\vx(t) = \vn(t)$, and hence the observed snapshots $\vx(t)$ that comprise $\mhC$ will consist entirely of noise.  In the setting where $\vx(t)$ is white Gaussian noise, Johansson~\cite{johansson2000sfa} derives the distribution of $\ell_1$, the principal eigenvalue of $\mhC$, for complex-valued data, and Johnstone~\cite{johnstone2001dle} gives the corresponding distribution in the real-valued case.  These results are defined in terms of the density function of the celebrated Tracy-Widom law (illustrated in Fig.~\ref{F:tw-density}) and imply the following theorem.
\begin{theorem}[Asymptotic Null Distribution]\label{T:null-value}
    Let $\ell_1 > 0$ be the largest eigenvalue of an $n$-dimensional sample covariance matrix $\mhC$ comprised of $N$ i.i.d.~observations $\vx$ according to~\eqref{E:C-empirical}, where each vector $\vx$ has i.i.d.~$\operatorname{Normal}(0,\sigma^2)$ entries. Defining the standardizing quantities
    \begin{align*}
        \mu_{N,n}
            &=
            \frac{\sigma^2}{N}
            \left( \sqrt{n} + \sqrt{N} \right)^2 \\
        \sigma_{N,n}
            &=
            \frac{\sigma^2}{N} \textstyle
            \left( \sqrt{n} + \sqrt{N} \right)
            \left( \frac{1}{\sqrt{n}} + \frac{1}{\sqrt{N}} \right)^{1/3},
    \end{align*}
    we obtain as $n,N\to\infty$, with $n/N\to \gamma$, the result
    \begin{equation*}
        \Prob \left\{ \frac{\ell_1 - \mu_{N,n}}{\sigma_{N,n}} \leq x \right\}
        \to
        F_\beta ( x ),
    \end{equation*}
    where $F_\beta( x )$ is the distribution function for the Tracy-Widom law of
    order $\beta$, with $\beta = 1$ if $\mhC \in \mathbb{R}^n$ and $\beta = 2$ if $\mhC \in \mathbb{C}^n$.
\end{theorem}

\begin{rem}
For complex $\mhC$, El Karoui~\cite{elkaroui2006rcr} obtains a convergence rate of order $(n\wedge N)^{2/3}$ through small modifications of $\mu_{N,n}$ and $\sigma_{N,n}$; Ma~\cite{ma2008atw} treats the case of real $\mhC$ similarly.
\end{rem}

Next, we consider properties of the empirical covariance estimator $\mhC$ when at least one signal is present, by way of the following sequence of nested alternate models:
\begin{equation}\label{E:SignalAlt}
    \left\{ \mathcal{M}_r\!:\, \vx(t) = \sum_{i=1}^{r} s_{i}(t) \va_{i} + \vn(t) \right\}, \quad 0 < r < n.
\end{equation}

In this setting, Baik et al.~\cite{baik2005pto} discovered a phase-transition phenomenon in the size of the population eigenvalue.  This work was further developed by Baik and Silversten~\cite{baik2006els}, Paul~\cite{paul2007ase}, and Onatski~\cite{onatski2007adp}.  Paul derived the limiting distribution for the case $\mhC \in \mathbb{R}^n$, with the ratio $n/N$ of sensors to snapshots tending to $\gamma < 1$.  Onatski later generalized this to $\gamma \in (0,\infty)$.  Finally, Bai and Yao~\cite{bai2008clt} derived these limiting distributions in the complex case.  A simplified summary of the above work is given by the following theorem.
\begin{theorem}[Asymptotic Alternate Distribution]\label{T:spiked-value}
    Consider a covariance $\mC = \E\left[ \vx(t) \vx(t)^* \right]$ under the model of~\eqref{E:SignalAlt}, with $r>0$ distinct principal eigenvalues $\{\lambda_i + \sigma^2\}$, and the corresponding
    sample covariance matrix $\mhC$, with $n$ ordered eigenvalues $\{\ell_i\}$.
    Denoting by $\Phi( \cdot )$ the standard Normal distribution function, and defining the standardizing quantities
    \begin{align*}
        \mu_{N,n}(\lambda)
            &= \left( \lambda + \sigma^2 \right)
                \left(1 + \gamma \sigma^2/\lambda \right) \\
        \sigma_{N,n}(\lambda)
            &= \left( \lambda + \sigma^2 \right) \textstyle
               \sqrt{
                   \frac{2}{\beta N}
                   \left( 1 - \gamma \sigma^4/\lambda^2 \right)
               },
    \end{align*}
    we have that if $\lambda_{1}, \lambda_{2}, \ldots, \lambda_{q} > \sqrt{\gamma} \sigma^2$, then
    \begin{equation*}
        \Prob \left\{ \frac{\ell_i - \mu_{N,n}(\lambda_i)}
                           {\sigma_{N,n}(\lambda_i)}       \leq x_i,
                            \quad
                            i = 1, \ldots, q \right\}
        \to
        \prod_{i=1}^q \Phi ( x_i ).
    \end{equation*}
    Otherwise, for any $\lambda_i\!:\! \lambda_i \leq \sqrt{\gamma} \sigma^2$, then $\ell_i \overset{a.s.}{\longrightarrow} \sigma^2 \left( 1 + \sqrt{\gamma} \right)^2$.
    As in Theorem~\ref{T:null-value}, $\beta = 1$ for real data and
    $2$ for complex data.
\end{theorem}
\begin{rem}
Theorem~\ref{T:spiked-value} yields a critical threshold $\sqrt{\gamma} \sigma^2$
below which any population eigenvalue is unrelated to its corresponding sample eigenvalue; sample eigenvalues corresponding to population eigenvalues above this threshold converge to a multivariate Normal with diagonal covariance.
\end{rem}

Paul and Onatski also give accompanying results linking the mutual information of population and corresponding sample eigen\emph{vectors} through the same critical threshold $\sqrt{\gamma} \sigma^2$, and moreover implying the general inconsistency of the latter as estimators of the former.  For the case of real-valued data, they prove the following theorem.

\begin{theorem}[Sample Eigenvector Inconsistency]\label{T:spiked-vector}
    Let $\vw_i$ denote the $i$th principal population eigenvector of $\E\left[ \vx(t) \vx(t)^T \right]$ under the model of~\eqref{E:SignalAlt}, and $\vhw_i$ its corresponding sample version via~\eqref{E:Cemp-decomp-eig}.  Then we have that
    \begin{equation*}
        \langle \vw_i, \vhw_j \rangle \overset{a.s.}{\longrightarrow}
        \begin{cases}
            \sqrt{\frac{\lambda_i - \gamma \sigma^4/\lambda_i}
                                   {\lambda_i + \gamma \sigma^2}} & \text{if $i = j$ and $\lambda_i > \sqrt{\gamma} \sigma^2$,} \\
            0 & \text{otherwise.}
        \end{cases}
    \end{equation*}
\end{theorem}
\begin{rem}
Onatski gives a convergence rate of $\sqrt{N}$ for the quantities of Theorem~\ref{T:spiked-vector}.
\end{rem}

\begin{figure}[t]
    \centering
    \includegraphics[width=\columnwidth]{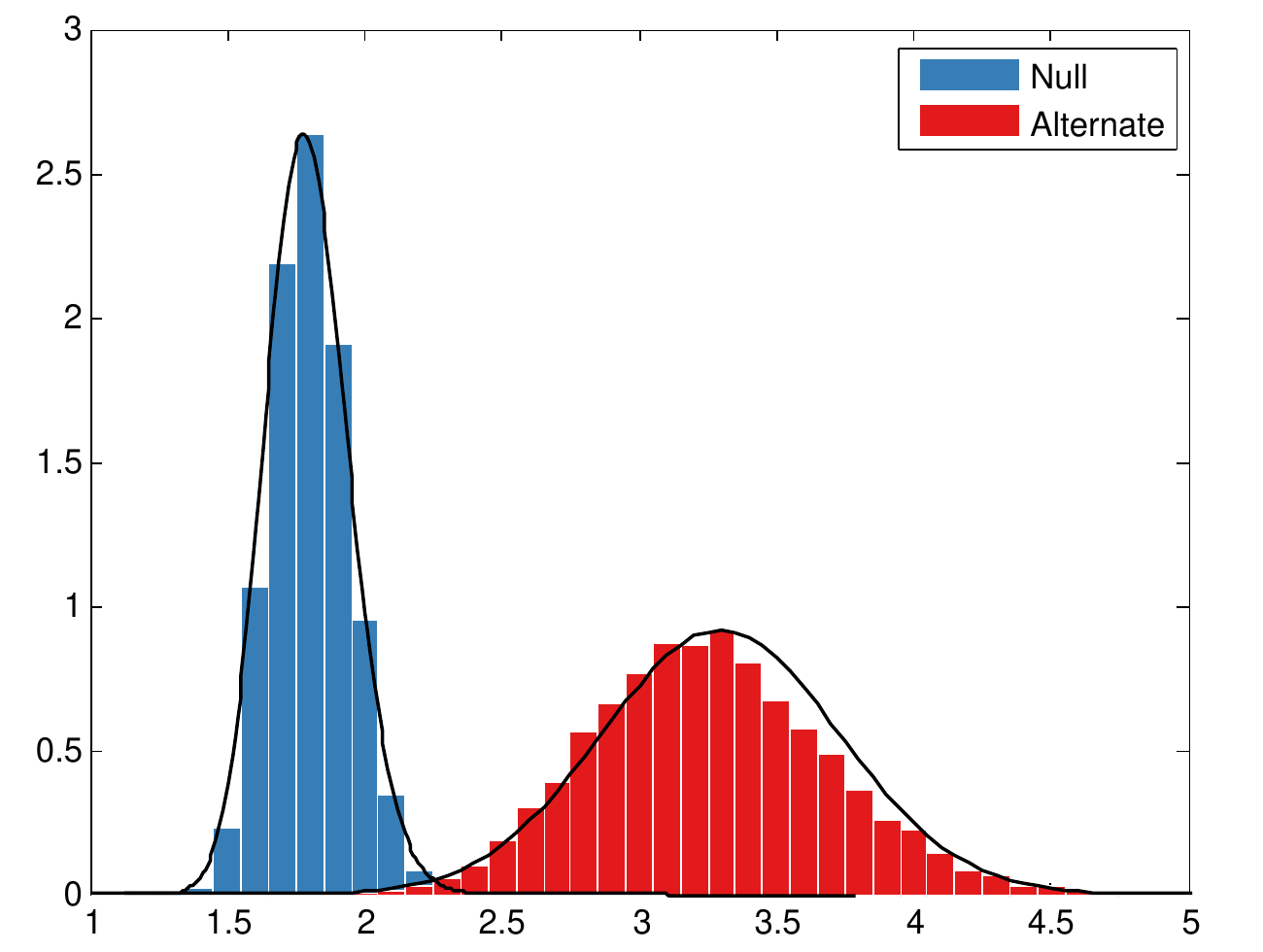}
    \caption{\label{F:null-alt-density}Example illustrating agreement of empirical and asymptotic distributions for the null and alternate settings of Theorems~\ref{T:null-value} and~\ref{T:spiked-value}, respectively.  The latter case comprises a single signal eigenvalue $\lambda$ set at a signal-to-noise ratio of 3~dB, with all other parameter settings matched to the simulation study of Section~\ref{S:emp-perf} (complex-valued data with $n=9$ and $N=45$).
    }
\end{figure}

To conclude this section, we note that while the above results are asymptotic in nature, evidence suggests that they are achieved in practice for small sample sizes.  In particular, Ma~\cite{ma2008atw} has catalogued empirical convergence rates for Theorem~\ref{T:null-value}, demonstrating that for $n$ ranging up to $500$, even with only $N=5$ samples the Tracy-Widom asymptotics remain a good approximation in the upper tail of the distribution---the setting of interest in the model selection problem posed here.  In later simulations, we apply our results to the model selection regime of direction-of-arrival estimation~\cite{kavcic1996are}, and consider complex-valued data in $n=9$ dimensions using $N=45$ snapshots.  Figure~\ref{F:null-alt-density} shows a comparison of empirical and asymptotic distributions in this scenario, with the generally good agreement providing further evidence for the practical utility of Theorems~\ref{T:null-value} and~\ref{T:spiked-value} above.

\section{Minimax-Optimal Rank Estimation}
\label{S:rank-est}

The previous section has given us a relatively complete description of the behavior of $\mhC$, our $n$-dimensional sample covariance comprised of $N$ array snapshots, with $n/N$ tending to $\gamma \in (0, \infty)$ as $n,N \to \infty$, and $\sigma^2$ the variance of additive white Gaussian noise in the observation model $\mathcal{M}_r$ of~\eqref{E:Signal}.  With this information, we are ready to proceed to the task of estimating the model order $r$, corresponding to the number of signals present.  In light of Theorems~\ref{T:spiked-value}~and~\ref{T:spiked-vector}, we need not consider the $r$th signal if its strength is below the critical threshold $\sqrt{\gamma} \sigma^2$, as the corresponding alternate model $\mathcal{M}_r$ will typically be indistinguishable from the null $\mathcal{M}_{r-1}$ in the asymptotic limit.  In the sequel we thus restrict our attention to the case when $\lambda_i > \sqrt{\gamma} \sigma^2$ for $i = 1, 2, \ldots, r$.

\subsection{Derivation of Asymptotic Risk for Signal Absence/Presence}

To formulate our minimax-optimal rank estimation task, we adopt a classical decision-theoretic approach: we first define a loss function $L$ to measure the quality of a particular estimate of $r$, and then derive a decision rule $\delta$ that minimizes the risk $R = \E \, [ L(\delta) ]$; i.e., the expected loss under our assumed probability model.

Consider first the most basic problem to which Theorems~\ref{T:null-value} and~\ref{T:spiked-value} offer a solution: differentiating between observing no signal at all ($r=0$) and observing a single signal ($r=1$).  When $r=0$, the snapshot $\vx(t)$ is assumed to be a zero-mean multivariate Normal with covariance $\sigma^2 \mI$. When the model order $r=1$, the population covariance matrix $\mC$ has one eigenvalue equal to $\lambda + \sigma^2$, and the rest equal to $\sigma^2$.  We encode these models by noting that $\lambda = 0$ in the first and $\lambda > 0$ in the second, and next address the task of choosing between them according to the sample covariance $\mhC$.

Using $\delta$ to denote a decision rule taking values in $\{ 0,1 \}$, we let $\delta = 0$ encode the decision $r = 0$. To this rule, we assign an ``inclusion'' penalty $\ci > 0$ for incorrectly overestimating the rank, and an ``exclusion'' penalty $\ce > 0$ for incorrectly underestimating it; when $\delta$ chooses the correct outcome we assign no penalty. We summarize this by introducing the loss function $L(\lambda, \delta)$, defined as
\begin{equation*}
    L(\lambda, \delta)
    =
    \begin{cases}
        \ci &\text{when $\lambda = 0$ and $\delta = 1$,} \\
        \ce &\text{when $\lambda > 0$ and $\delta = 0$,} \\
        0   &\text{otherwise.}
    \end{cases}
\end{equation*}

Guided by the results of Section~\ref{S:null-alt-rmt}, we distinguish between the two cases above based on $\ell_1(\mhC)$, the principal eigenvalue of the observed sample covariance matrix $\mhC$: If $\ell_1$ is larger than some fixed threshold, we estimate $r$ as $\hat r = 1$, and otherwise we choose $\hat r = 0$.  For a threshold $T$, we thus define our decision rule $\delta$ as
\begin{equation}\label{E:delta}
    \delta_T (\ell)
    =
    \begin{cases}
        1 &\text{if $\ell > T$,} \\
        0 &\text{otherwise.}
    \end{cases}
\end{equation}

The risk associated with this rule is given by evaluating the expected loss $\E \, [ L( \lambda, \delta_T( \ell_1 ) )]$ associated with our chosen test statistic $\ell_1(\mhC)$, with respect to probabilities $\Prob_{\mathcal{M}_i}$ under the two competing models $\mathcal{M}_0$ and $\mathcal{M}_1$:
\begin{equation*}
    R( \lambda, \delta_T )
         = \begin{cases}
               \ci \cdot \Prob_{\mathcal{M}_0} \! \left\{ \ell_1 > T \right\}
                   &\text{when $\lambda=0$,} \\
               \ce \! \cdot \Prob_{\mathcal{M}_1} \! \left\{ \ell_1 \leq T \right\}
                   &\text{otherwise.}
           \end{cases}
\end{equation*}
Theorem~\ref{T:null-value} in turn describes the asymptotic distribution of $\ell_1$ when
$r=0$, while Theorem~\ref{T:spiked-value} describes it when $r=1$.  We thus obtain a precise asymptotic description of the risk $R$ as
\begin{equation}\label{E:asympt-risk}
    R( \lambda, \delta_T )
    \to \begin{cases}
           \ci \cdot \left(1 - F_\beta \left( \frac{ T - \mu_{N,n} }
                                                   { \sigma_{N,n}} \right) \right)
               &\text{when $\lambda=0$,} \\
           \ce \cdot \Phi \left( \frac{T - \mu_{N,n} (\lambda)}
                                      { \sigma_{N,n} (\lambda)} \right)
               &\text{otherwise,}
       \end{cases}
\end{equation}
where again $\Phi( \cdot )$ denotes the standard normal CDF.

Suppose we have knowledge that when a signal is present, its strength is at least equal to $\ulambda$.  We may then choose a threshold $T$ to minimize the maximum risk over all relevant scenarios.  Specifically,
we seek to minimize
\begin{equation}\label{E:mm-risk}
    \sup_{\lambda \in \{ 0 \} \cup [\ulambda, \infty)}  \!\!\!\! R( \lambda, \delta_T )
    =
     R(0, \delta_T) \vee R(\ulambda, \delta_T)
    .
\end{equation}
It is not hard to show that this occurs when
\(
    R(0, \delta_T) = R(\ulambda, \delta_T),
\)
and hence we conclude from~\eqref{E:asympt-risk} that $T$ must solve
\begin{equation}\label{E:T-eqn}
    \textstyle
    \ci \cdot \left(1 - F_\beta \left( \frac{ T - \mu_{N,n} }
                                         { \sigma_{N,n}} \right) \!\right)
   \! =
    \ce \cdot \Phi \left( \frac{T - \mu_{N,n} (\ulambda)}
                            { \sigma_{N,n} (\ulambda)} \right) \!.
\end{equation}

\subsection{Asymptotic Analysis of Minimax Threshold Behavior}

While it is easy to compute the minimax-optimal $T$ in~\eqref{E:T-eqn} numerically using bisection, and therefore implement the decision rule $\delta_T( \ell_1 )$ of~\eqref{E:delta}, we know of no closed-form expression for $T$.  Instead, we now present a brief asymptotic analysis of the minimax threshold behavior in order to gain insight as to how $T$ behaves as $\ulambda$ varies.  For clarity of presentation and without loss of generality, we assume $\sigma^2 = 1$ in the sequel.

Our first observation stems from a comparison of the mean standardization quantities in Theorems~\ref{T:null-value} and~\ref{T:spiked-value}: It is easily verified that $\mu_{N,n}(\lambda) \to \mu_{N,n}$ as
$\lambda \to \sqrt{\gamma}$, implying a need for analysis when
the minimal assumed signal strength $\ulambda$ is close to $\sqrt{\gamma}$.
Indeed, for other values of $\ulambda$ a threshold $T$ slightly above $\mu_{N,n}$ will yield minimax risk very close to $0$, since $\sigma_{N,n} \sim N^{-2/3}$ and $\sigma_{N,n}(\ulambda) \sim N^{-1/2}$ in this case.

To study the variation of $T$ with $\ulambda$ in this regime, we first parameterize $\ulambda$ in $h$, with the restriction $h > 0$, as
\begin{equation}\label{E:Lh}
    \ulambda(h) = \sqrt{ \gamma } + h.
\end{equation}
The threshold behavior of interest occurs when $h$ is of size $N^{-1/3}$, and so we parametrize $T$ in $t$ for this case as
\begin{equation}\label{E:Tt}
    T(t) = \mu_{N,n} + t \sigma_{N,n}.
\end{equation}
We summarize the behavior of $t$ for $h$ near $N^{-1/3}$ in the following two lemmas, whose proofs are given in Appendix~\ref{S:mm-asymp}.

\begin{lemma}\label{L:thresh-small-h}
    Let  $\ulambda(h)$ and $T(t)$ be parameterized as in~\eqref{E:Lh} and~\eqref{E:Tt}, respectively, and fix $h = o\left(N^{-1/3}\right)$.  The behavior of $t$ then depends on the ratio of costs $\ce$ and $\ci$ as follows:
    \begin{enumerate}
    \item If $\ce > (1 -  F_\beta(0)) \cdot \ci$, then
    \begin{multline*}
        t
        =
        2
        \beta^{-1/2}
        \left(
            \frac{h^3 N}{ \gamma^{1/4} + \gamma^{-1/4} }
        \right)^{1/6}
        \\
        \cdot
        \Phi^{-1} \! \left(
            \frac{\ci}{\ce}
            \left(
                1 - F_\beta(0)
            \right)
        \right)
        \left(
            1 + o(1)
        \right).
    \end{multline*}
    \item If $\ce < (1 -  F_\beta(0)) \cdot \ci$, then for
    \(
        t_{\sqrt{\gamma}}
        =
        F_\beta^{-1} \!\left( 1 - \ce/\ci \right),
    \)
    we have that
    \begin{multline*}
        t
        =
        t_{\sqrt{\gamma}}
        +
        \left[
            f_\beta \left( t_{\sqrt{\gamma}} \right)
        \right]^{-1}
        \cdot
        \frac{\ce}{\ci}
        \sqrt{\frac{2}{\beta \pi}}
        \left(
            \frac{ h^3 N }{\gamma^{1/4} + \gamma^{-1/4}}
        \right)^{1/6}
        \\
        \cdot
        t_{\sqrt{\gamma}}^{-1} \,
        \exp\!\left(
            -
            \frac{\beta t_{\sqrt{\gamma}}^2}{8}
            \left(
                \frac{\gamma^{1/4} + \gamma^{-1/4}}{ h^3 N }
            \right)^{1/3}
        \right)
        \left(
            1 + o(1).
        \right)
    \end{multline*}
    \item If $\ce = (1 -  F_\beta(0)) \cdot \ci$, then $t$ solves
    \begin{multline*}
        t^2
        =
        \left[
               f_\beta ( 0 )
        \right]^{-1}
        \cdot
        \frac{\ce}{\ci}
        \sqrt{\frac{2}{\beta \pi}}
        \left(
            \frac{ h^3 N }{\gamma^{1/4} + \gamma^{-1/4}}
        \right)^{1/6}
        \\
        \cdot
        \exp \!\left(
            -
            \frac{\beta t^2}{8}
            \left(
                \frac{\gamma^{1/4} + \gamma^{-1/4}}{ h^3 N }
            \right)^{1/3}
        \right)
        \left(
            1 + o\left( 1 \right)
        \right).
    \end{multline*}
    \end{enumerate}
\end{lemma}

\begin{lemma}\label{L:thresh-med-h}
    Suppose instead that $h = h_0 N^{-1/3}$ for some constant $h_0 > 0$.  Then we have the result that
    \begin{multline*}
        \ci
        \cdot
        (1 - F_\beta ( t ) )
        \sim
        \ce
        \cdot
        \Phi \!\left(
            \frac{\beta^{1/2} t}{ 2 \sqrt{ h_0 }}
            \left( \gamma^{1/4} + \gamma^{-1/4} \right)^{1/6} \right.
            \\
            -
            \left.
            \frac{ \beta^{1/2} h_0^{3/2} }
                 { \sqrt{ 2 \gamma }
                   \left( \gamma^{1/4} + \gamma^{-1/4} \right) }
        \right).
    \end{multline*}
    Moreover, if it is also the case that $\ce = \omega( \ci )$, then
    \[
        t
        \sim
        \,-
        \sqrt{
            \frac{ 8 h_0 }{ \beta (\gamma^{1/4} + \gamma^{-1/4} )^{1/6} }
            \log \frac{\ce}{\ci}
        } \,;
    \]
    if instead we have that $\ce = o( \ci )$, then
    $
        t
        \sim
        \left(
            \frac{3}{2 \beta}
            \log \frac{\ci}{\ce}
        \right)^{2/3}.
    $
\end{lemma}

\subsection{Extension to the General Model Order Selection Problem}

In the above discussion we treated the basic model order selection problem of $\mathcal{M}_r: r = 0$ versus $\mathcal{M}_r: r = 1$, in order to develop our problem formulation and asymptotic results.  In practice, of course, techniques are needed to address the general model order selection problem of estimating $r \geq 0$.  The main result of this section is that an asymptotically minimax-optimal rank selection rule is in fact obtained through repeated application of the basic $\mathcal{M}_0$ vs.~$\mathcal{M}_1$ case.

To develop such a rule and verify its properties, we first extend the thresholding approach seen earlier to the case of arbitrary $r > 0$.  Rather than specifying only a single cost $\ce$ for incorrectly excluding a term, we now require a sequence $\{\ce(i)\}_{i=1}^n$  of nonnegative costs, corresponding to exclusion of respective signal terms.  To this end we define a cumulative exclusion cost $\Ce(\cdot)$ as
\begin{equation}\label{E:Ce}
    \Ce(j) = \sum_{i=j}^n \ce(i), \quad 1 \leq j \leq n \text{.}
\end{equation}
One possible choice for the sequence $\{\ce(i)\}_{i=1}^n$ is simply to set all exclusion costs to be equal; alternatively, with prior knowledge that there are at most $r_\text{max}$ signals, one might well set $\ce(i) = 0$ for
$i > r_\text{max}$.

In a similar manner, we define a sequence of thresholds $\{T(i)\}_{i=1}^n$ associated to the sequence of ordered sample eigenvalues $\{\ell_i\}_{i=1}^n$, with each threshold $T(i)$ determined by an inclusion cost $\ci$ and the corresponding exclusion cost $\Ce(i)$.  An ordered rank selection procedure for $r$ is then given by Algorithm~\ref{A:rank-selection} below, and verified to be asymptotically minimax optimal by the theorem that follows.
\begin{algorithm}
\caption{\label{A:rank-selection}Minimax-optimal rank selection procedure}
\begin{enumerate}
\item Fix a threshold sequence $\{T(i)\}_{i=1}^n$ via~\eqref{E:T-eqn} and pre-assigned erroneous inclusion/exclusion costs $\ci, \Ce(i)$;
\item Form the sample covariance $\mhC$, set $i \leftarrow 1$, and test:

    \begin{algorithmic}
    \WHILE{ $\ell_{i}(\mhC) > T( i )$ }
        \STATE $i \leftarrow i + 1$
    \ENDWHILE
\end{algorithmic}
\item Return $\hat r \leftarrow i - 1$ as the final estimate of rank $r$.
\end{enumerate}
\end{algorithm}

\begin{theorem}[Minimax Optimality]\label{T:mm-opt} For fixed, nonnegative inclusion cost $\ci$ and exclusion
costs $\{\Ce(i)\}$, the rank selection procedure of Algorithm~\ref{A:rank-selection} is asymptotically minimax.
\end{theorem}

\begin{IEEEproof}
Note that the $i$th iteration of Step~2 in Algorithm~\ref{A:rank-selection} tests $\mathcal{M}_r: r = i-1$ versus $\mathcal{M}_r: r \geq i$.  By~\eqref{E:mm-risk}, the worst-case risk occurs when $r \to \infty$ and each signal has strength equal to the minimum assumed strength $\ulambda$, in which case the cost for excluding them all is given by $\Ce(i)$ in~\eqref{E:Ce}. Thus the optimal threshold at this stage is given by the corresponding $T(i)$ satisfying~\eqref{E:T-eqn}.  Since Theorem~\ref{T:spiked-value} proves the asymptotic independence of the $r$ principal sample eigenvalues, we conclude in turn that considering only the $i$th eigenvalue $\ell_i$ at the $i$th iteration incurs no loss in power.
\end{IEEEproof}

According to~\eqref{E:T-eqn}, knowledge of the noise power $\sigma^2$ is required to determine optimal thresholds.  In an adaptive setting, we note that $\sigma^2$ may be estimated at time $t$ by way of the residual variance from the $\hat r(t-1)$ signals at time index $t-1$~\cite{rabideau1996fra}.  In a non-adaptive setting, Kritchman and Nadler~\cite{kritchman2008dnc} and Patterson et al.~\cite{patterson2006psa} suggest alternative approaches.

\section{Empirical Performance Examples}
\label{S:emp-perf}

\begin{figure*}
    \includegraphics[width=2.1\columnwidth]{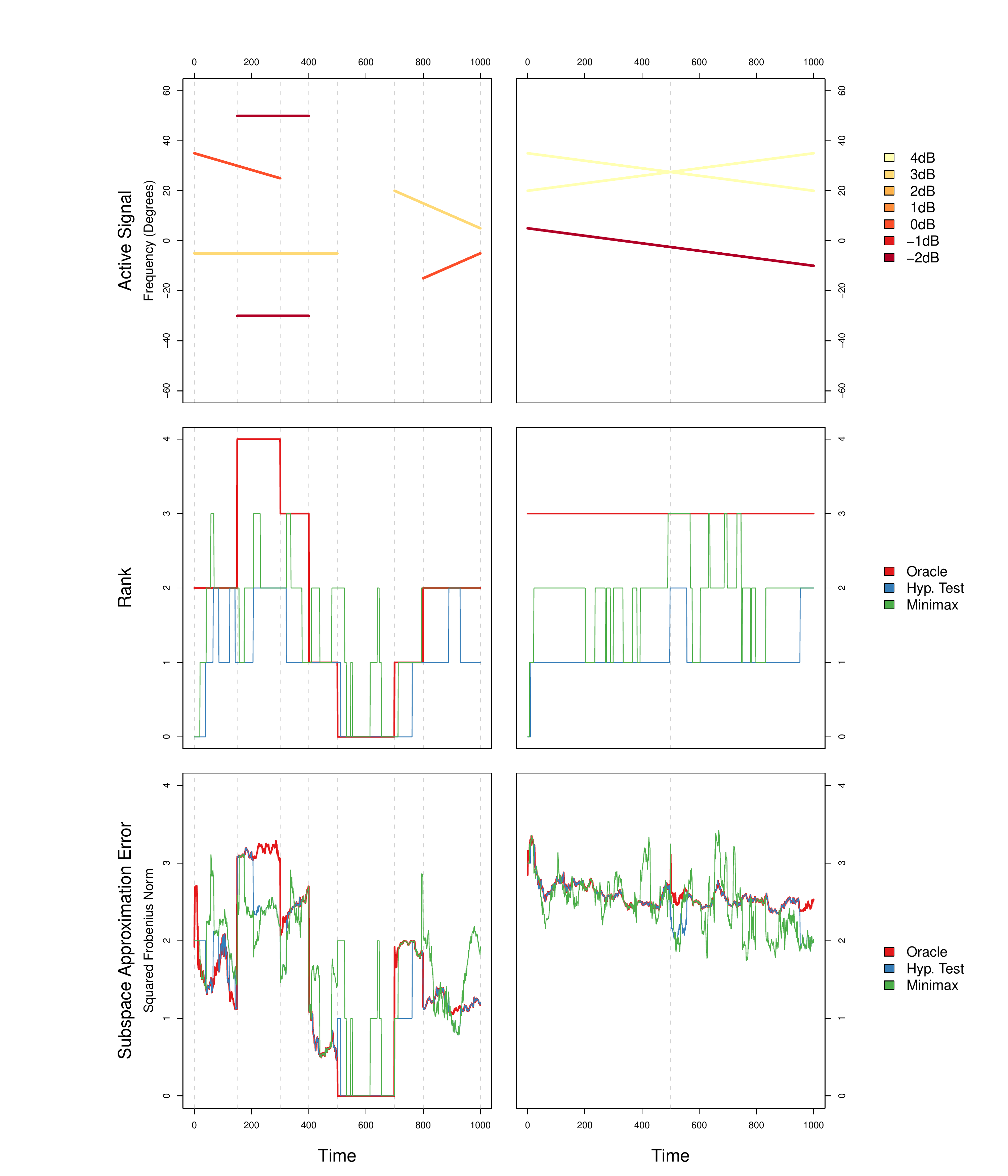}
    \caption{\label{F:simulations} Two simulations taken from a direction-of-arrival estimation setting, showing empirical performance of the minimax rank selection rule of Algorithm~\ref{A:rank-selection} and the hypothesis-testing-based rule of~\cite{kritchman2008dnc}.  Signals of the form $C_0 ( 1, e^{j \omega}, \ldots, e^{j 8 \omega})$ appear and disappear over time $t \in [0,1000]$, with signal directions and strengths shown in the top row.  One snapshot is sampled per unit time, and the subspace at time $t$ is estimated using the snapshots for times in the window $(t-45,t]$.  The second row of the plot shows the true rank $r$ as a function of $t$, along with the estimated rank $\hat r$ according to the method of Algorithm~\ref{A:rank-selection}.  The subspace approximation error in squared Frobenius norm, $\|\mW \mW^\ast - \mhW_k \mhW^\ast_k \|_F^2$, is plotted for $k = r, \hat r$ in the third row.}
\end{figure*}

Having derived a rank selection rule in the preceding section and investigated its theoretical properties, we now provide two brief simulation studies designed to demonstrate the empirical performance of this procedure.  We report on an evaluation of Algorithm~\ref{A:rank-selection} by way of two simulations adopted from the direction-of-arrival estimation setting of~\cite{kavcic1996are}, shown in the top panels of Fig.~\ref{F:simulations}.  The first simulation has signals of different strengths appearing and disappearing over time, leading to a varying rank, whereas the second simulation comprises a constant number of signals.  In both settings, the snapshots are in $n = 9$ dimensions, and arrival directions vary over time.

Each simulation features a range of signal strengths, including some so low as to be indistinguishable from the background noise.  In the first simulation, for example, there is one signal below the detection threshold $\sqrt{\gamma}$ between time indices $150$ and $400$; in the second simulation, the weakest signal is always below the detection threshold, implying that it will not in general be detected.

In these performance examples we employed a windowed covariance estimate with $N = 45$ observations, and set $\ci = \ce(1) = \cdots = \ce(n)$, with $\lambda_0 = \sqrt{\gamma} + N^{-1/3}$.  We treat $\sigma^2$ as unknown, and estimate it for every time $t$ as the mean of the estimated noise eigenvalues at the previous time step $t-1$~\cite{rabideau1996fra}.  The middle panels of Fig.~\ref{F:simulations} demonstrate the corresponding rank estimation results, from which we observe good empirical agreement with theoretical predictions; where we incur estimation errors, they tend to be as a result of signals whose strength falls below $\sqrt{\gamma}$, the detection limit.  The bottom panels of Fig.~\ref{F:simulations} also show the error in squared Frobenius norm, $\| \mW \mW^\ast - \mhW_k \mhW^\ast_k \|_F^2$, of the corresponding subspace estimate
\(
    \mhW_k
    =
    \begin{pmatrix}
        \vhw_1 & \vhw_2 & \cdots & \vhw_k
    \end{pmatrix}
\)
for $k = r$ and $k = \hat r$, respectively the true and estimated ranks.

\begin{figure*}
    \centering
    \includegraphics[width=2.1\columnwidth]{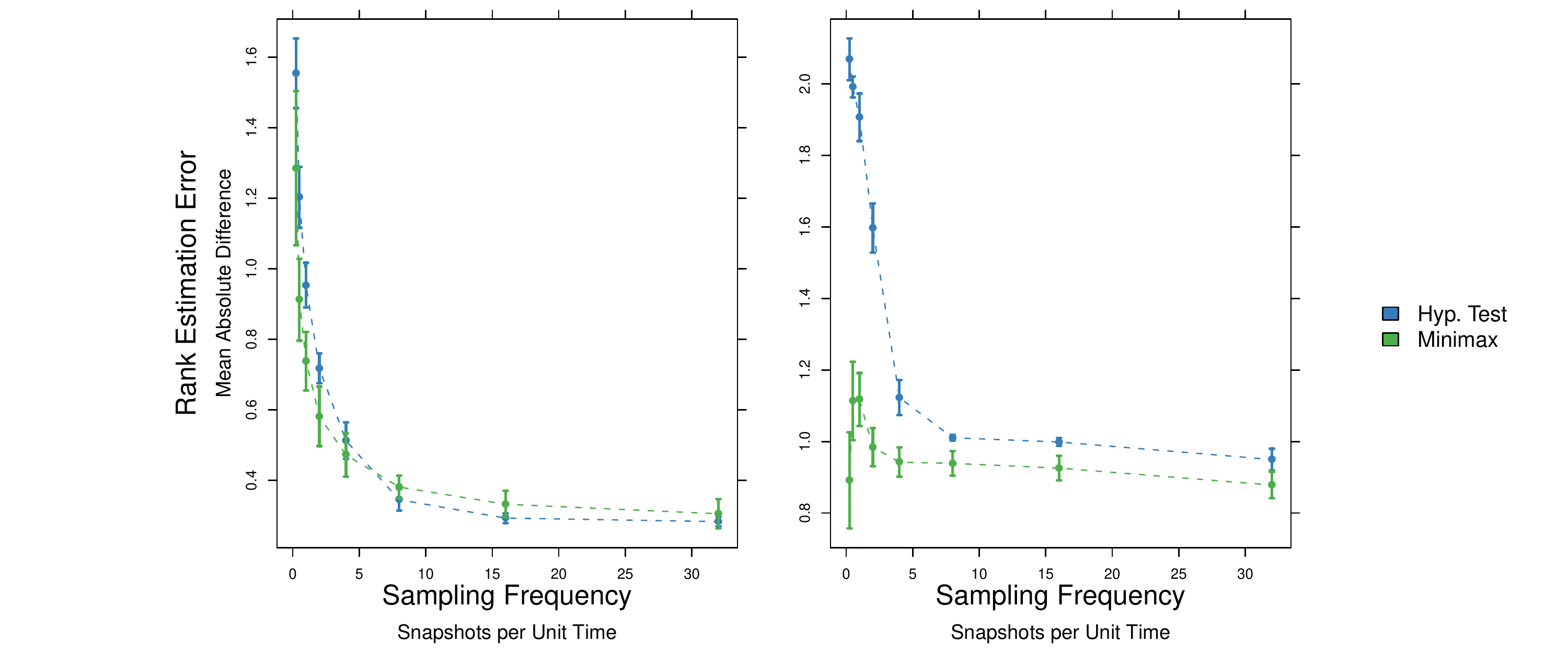}
    \caption{Simulations demonstrating that the performance of the rank estimators improves as the snapshot sampling frequency increases.  With the same signal arrival patterns as in Fig.~\ref{F:simulations}, we vary the snapshot sampling rate and show how the rank estimation error $\frac{1}{1000} \int_0^{1000} |r(t) - \hat r(t)| \, dt$ behaves for the two rank estimators $\hat r$.  The points show the mean behavior and the error bars show $1$ standard deviation, computed from $50$ replicates of the experiment.}\label{F:consistency-sim}
  \end{figure*}

For purposes of comparison, we also show a hypothesis-testing-based approach adopted from Kritchman and Nadler~\cite{kritchman2008dnc}, based on Tracy-Widom quantiles and set at a fixed 0.5\% false alarm rate as suggested by those authors.  We note however, that the variance estimation approach of~\cite{kritchman2008dnc} does not apply directly in the subspace tracking scenario, as it employs all eigenvalues of the sample covariance matrix; here we employed an estimate based on the residual from the previous time step.  We see that in comparison to the proposed method of Algorithm~\ref{A:rank-selection}, this approach tends to provide a consistently more conservative estimate of rank, resulting in greater overall error in this simulation context.

In Fig.~\ref{F:consistency-sim}, we show how these rank estimators perform as the snapshot sampling frequency per unit time is increased.  Since the number of sensors is held fixed, this corresponds to varying $\gamma$; the simulations of Fig.~\ref{F:consistency-sim} demonstrate that for higher sampling frequencies (lower values of $\gamma$), the rank estimation problem becomes easier and the error decreases.  Note in the left-hand panel of Fig.~\ref{F:consistency-sim}, however, that because the rank is not constant within all time windows, the resultant error will not necessarily asymptote to zero.  In contrast, that of the right-hand panel \emph{will} eventually reach zero when $\gamma$ becomes sufficiently small to render the weakest signal detectable (as per Theorem~\ref{T:spiked-value}).

\section{Discussion}
\label{s:summ}

In this article we have presented sample covariance asymptotics stemming from random matrix theory, and have in turn brought them to bear on the problem of optimal rank estimation.  This task poses a classical model order selection problem that arises in a variety of important statistical signal and array processing systems, yet is addressed relatively infrequently in the extant literature.  Key to our approach is the existence of a phase transition threshold in the context of the standard array observation model with additive white Gaussian noise, below which eigenvalues and associated eigenvectors of the sample covariance fail to provide any information on population eigenvalues.  Using this and other results, we then developed a decision-theoretic rank estimation framework that led to a simple ordered selection rule based on thresholding; in contrast to competing approaches, this algorithm was shown to admit asymptotic minimax optimality and to be free of tuning parameters.  We concluded with a brief simulation study to demonstrate the practical efficacy of our rank selection procedure, and plan to address a diverse set of rank estimation tasks as part of our future work.

\appendices

\section{Tracy-Widom Asymptotics}\label{S:tw-asymp}

A characterization of the tail behavior of $F_\beta(s)$, the distribution function of the Tracy-Widom law, is required to obtain the results of Lemmas~\ref{L:thresh-small-h} and~\ref{L:thresh-med-h}.  In this appendix we derive the asymptotic properties of $F_\beta(s)$ for $\beta=1,2$ as
$|s| \to \infty$.

To begin, let $q(x)$ solve the Painlev\'e~II equation
\[
    q''(x) = x q(x) + 2 q^3(x),
\]
with boundary condition $q(x) \sim \Ai(x)$ as $x \to \infty$ and $\Ai(x)$ the Airy function. Then it follows that
\begin{align*}
    F_1(s)
        &= \exp \left\{
            -\half \int_s^\infty q(x) + (x - s) q^2(x) dx
        \right\}, \\
\intertext{and}
    F_2(s)
        &= \exp \left\{
            - \int_s^\infty (x - s) q^2(x) dx
        \right\}.
\end{align*}
As $x \to \infty$, the Airy function behaves as
\(
    \Ai(x)
        \sim \frac{1}{2\sqrt{\pi}} x^{-1/4}
             \exp \left( -\frac{2}{3} x^{3/2} \right);
\)
asymptotic properties of $q$ as $x \to -\infty$ are studied by Hastings and McLeod \cite{hastings1980bvp}, who show that in this case,
\(
    q(x) \sim \sqrt{ |x| / 2 }.
\)
Using these facts, we can compute for $s \to \infty$ the term
\begin{align*}
    \int_s^\infty & q(x) dx \\
        & \sim \frac{1}{2 \sqrt{\pi}}
              \int_s^\infty
                   x^{-1/4} \exp\left( -\frac{2}{3} x^{3/2} \right) dx \\
        &= \frac{1}{2 \sqrt{\pi}}
           \int_0^\infty
               \exp\left(
                   -\frac{2}{3} (x + s)^{3/2} - \frac{1}{4} \log (x + s)
               \right) dx \\
        &\sim \frac{1}{2 \sqrt{\pi}}
              \int_0^\infty
                  \exp\left(
                      -\frac{2}{3} s^{3/2}
                      \left(
                          1 + \frac{3}{2} \frac{x}{s}
                      \right)
                      - \frac{1}{4} \log s
                  \right)
                  dx \\
        &= \frac{1}{2 \sqrt{\pi}} s^{-1/4}
           \exp\left( -\frac{2}{3} s^{3/2} \right)
           \int_0^\infty \exp\left( -s^{1/2} x \right) dx \\
        &= \frac{1}{2 \sqrt{\pi}} s^{-3/4}
           \exp\left( -\frac{2}{3} s^{3/2} \right) ,
\end{align*}
and also
\begin{align*}
    \int_s^\infty & (x - s) q^2(x) dx \\
        & \sim \frac{1}{4 \pi}
              \int_s^\infty
                  (x - s)
                  x^{-1/2}
                  \exp\left( -\frac{4}{3} x^{3/2} \right)
                  dx \\
        &= \frac{1}{4 \pi}
           \int_0^\infty
               x
               \exp\left(
                   -\frac{4}{3} (x + s)^{3/2}
                   -\frac{1}{2} \log( x + s )
               \right)
               dx \\
        &\sim \frac{1}{4 \pi}
              \int_0^\infty
              x
              \exp\left(
                  -\frac{4}{3}s^{3/2}
                  \left( 1 + \frac{3}{2} \frac{x}{s} \right)
                  - \frac{1}{2} \log s
              \right)
              dx \\
        &= \frac{1}{4 \pi}
           s^{-1/2}
           \exp\left( -\frac{4}{3} s^{3/2} \right)
           \int_0^\infty
                x
                \exp\left(
                    -2 s^{1/2} x
                \right)
                dx \\
        &= \frac{1}{16 \pi}
           s^{-3/2}
           \exp\left( -\frac{4}{3} s^{3/2} \right).
\end{align*}
Likewise, for $s \to -\infty$ we have that
\begin{align*}
    \int_s^\infty q(x) dx &\sim \frac{\sqrt{2}}{3} |s|^{3/2}, \\
\intertext{and}
    \int_s^\infty (x - s) q^2(x) dx &\sim \frac{|s|^3}{12}.
\end{align*}
Now, we must have that as $s\to\infty$,
\begin{align*}
    F_1(s)
        &\sim \exp\left\{
                  -\frac{1}{2}
                  \left(
                      \frac{1}{2 \sqrt{\pi}}
                      s^{-3/4}
                      e^{-\frac{2}{3} s^{3/2}}
                      \right. \right.
                      \\ & \qquad \qquad \qquad \quad\,
                      \left. \left.
                      +
                      \frac{1}{16 \pi}
                      s^{-3/2}
                      e^{-\frac{4}{3} s^{3/2}}
                  \right)
              \right\} \\
        &\sim \exp\left\{
                   -
                   \frac{1}{4 \sqrt{\pi}}
                   s^{-3/4}
                   e^{-\frac{2}{3} s^{3/2}}
              \right\} \\
        &\sim 1
              -
              \frac{1}{4 \sqrt{\pi}}
              s^{-3/4}
              \exp\left( -\frac{2}{3} s^{3/2} \right), \\
\intertext{and similarly}
    F_2(s) &\sim 1
                 -
                 \frac{1}{16 \pi}
                 s^{-3/2}
                 \exp\left( -\frac{4}{3} s^{3/2} \right).
\end{align*}
We also get that as $s\to -\infty$,
\begin{align*}
    F_1(s) &\sim \exp\left( -\frac{|s|^3}{24} \right),
\intertext{and}
    F_2(s) &\sim \exp\left( -\frac{|s|^3}{12} \right).
\end{align*}
In summary, then, for $\beta = 1,2$ we have that as $s \to -\infty$,
\begin{equation}\label{E:tw-s-neg}
    F_\beta(s) \sim \exp\left( -\frac{\beta}{24} |s|^3 \right),
\end{equation}
while for $s \to \infty$ we have
\begin{equation}\label{E:tw-s-pos}
    1 - F_\beta(s)
    \sim
    \left(
        \frac{1}{16 \pi}
    \right)^{\beta/2}
    s^{-3 \beta/4 }
    \exp\left(
        -
        \frac{2 \beta}{3}
        s^{3/2}
    \right).
\end{equation}

\section{Minimax Threshold Asymptotics}\label{S:mm-asymp}

In this appendix we prove Lemmas~\ref{L:thresh-small-h} and~\ref{L:thresh-med-h}.  Recall that by~\eqref{E:Lh} we have the parameterization $\ulambda(h) = \sqrt{\gamma} + h$ for some fixed $h > 0$, and we seek asymptotic properties of the associated minimax eigenvalue threshold $T$, parameterized according to~\eqref{E:Tt} as $T(t) = \mu_{N,n} + t \sigma_{N,n}$.  To derive the asymptotic behavior of $T$ for small and large $h$, we first require the asymptotic
behaviors of $\mu_{N,n}( \sqrt{\gamma} + h )$ and $\sigma_{N,n}(\sqrt{\gamma} + h)$ for small $h$, along with the tail behaviors of $\Phi(x)$ and $F_\beta(x)$.

To this end, it is not hard to show that for small $h$,
\begin{align*}
    \mu_{N,n}(\sqrt{\gamma} + h)
        &= \mu_{N,n} + \frac{h^2}{\sqrt{\gamma}} + O(h^3), \\
    \sigma_{N,n}(\sqrt{\gamma} + h)
        &= 2 \beta^{-1/2}
            (\gamma^{1/4} + \gamma^{-1/4}) \sqrt{\frac{h}{N}}
            + O \left( \frac{h}{\sqrt{N}} \right).
\end{align*}
Since
\(
    \sigma_{N,n}
    =
    \left(
        (\gamma^{1/4} + \gamma^{-1/4})/N
    \right)^{2/3},
\)
for $h = O(N^{-1/2})$ we have that
\begin{multline*}
    \frac{ T - \mu_{N,n}(\sqrt{\gamma} + h)}
         { \sigma_{N,n}(\sqrt{\gamma} + h) }
    =
    \beta^{1/2}
    \frac{t}{2}
    \left(
        \frac{\gamma^{1/4} + \gamma^{-1/4}}{ h^3 N }
    \right)^{1/6}
    \\
    -
    \sqrt{\frac{\beta}{2 \gamma}}
    \cdot
    \frac{\sqrt{ h^3 N }}{\gamma^{1/4} + \gamma^{-1/4}}
    + O \left( h/\sqrt{N} \right) .
\end{multline*}

Using the notation $x \sim y$ to denote $x = y\left( 1 + o(1) \right)$, a
standard result \cite{abramowitz1970hmf} is that as $x \to \infty$, we have
\[
    1 - \Phi(x)
    \sim
    \frac{1}{\sqrt{2 \pi}}
    x^{-1}
    \exp\left( -\frac{1}{2} x^2 \right).
\]
Therefore, as $\epsilon \to 0$,
\(
    \Phi^{-1} (\epsilon)
    \sim
    -
    \sqrt{ 2 \log \epsilon^{-1} }
\).

Tail properties of $F_\beta(x)$ are derived in Appendix~\ref{S:tw-asymp}, and given by~\eqref{E:tw-s-neg} and~\eqref{E:tw-s-pos}.  From these, we have that as $\epsilon \to 0$,
\[
    F_\beta^{-1} ( \epsilon )
    \sim
    -
    \left( 24 \beta^{-1} \log \epsilon^{-1} \right)^{1/3}
\]
and
\[
    F_\beta^{-1} ( 1 - \epsilon )
    \sim
    \left(
        \frac{3}{2 \beta}
        \log \epsilon^{-1}
    \right)^{2/3}.
\]

Equipped with these results, we are now ready to give the proofs of Lemmas~\ref{L:thresh-small-h} and~\ref{L:thresh-med-h}.
\begin{IEEEproof}[Proof of Lemma~\ref{L:thresh-small-h}]
If $h = o\left(N^{-1/3}\right)$ then
\begin{multline*}
    \Phi \!
    \left(
        \frac{ T - \mu_{N,n}(\sqrt{\gamma} + h)}
             { \sigma_{N,n}(\sqrt{\gamma} + h) }
    \right)
    =
    \Phi \!
    \left(
    \frac{\beta^{1/2}t}{2}
    \left(
        \frac{\gamma^{1/4} + \gamma^{-1/4}}{ h^3 N }
    \right)^{1/6}
    \right)
    \\
    +
    O\left(\sqrt{h^3 N} \right)
    .
\end{multline*}
When $t = O\left((h^3 N)^{1/6}\right)$, the left-hand side of~\eqref{E:T-eqn} converges to $\ci ( 1 - F_\beta(0))$.  Therefore,
\begin{multline*}
    t
    =
    2
    \beta^{-1/2}
    \left(
        \frac{ h^3 N }{\gamma^{1/4} + \gamma^{-1/4}}
    \right)^{1/6}
    \Phi^{-1}
    \left(
        \frac{\ci}{\ce}
        \Big( 1 - F_\beta(0) \Big)
    \right)
    \\
    +
    O\left( (h^3 N)^{1/3} \right)
    .
\end{multline*}
Of course, this only makes sense when $\ce > (1 - F_\beta(0)) \cdot \ci$.
Otherwise, we must have $t = \omega\left((h^3 N)^{1/6} \right)$.  In this
case, using $\Phi(x) = 1 - (1/\sqrt{2 \pi}) x^{-1} \exp(-x^2/2)(1 + O(x^{-2}))$
as $x \to \infty$, we obtain for $t > 0$ the expression
\begin{multline*}
    \Phi
    \left(
        \frac{ T - \mu_{N,n}(\sqrt{\gamma} + h)}
             { \sigma_{N,n}(\sqrt{\gamma} + h) }
    \right)
    =
    1
    -
    \sqrt{\frac{2}{\beta \pi}}
    \left(
        \frac{ h^3 N }{\gamma^{1/4} + \gamma^{-1/4}}
    \right)^{1/6}
    \\
    \cdot t^{-1}
    \exp
    \left(
        -\frac{\beta t^2}{8}
        \left(
            \frac{\gamma^{1/4} + \gamma^{-1/4}}{ h^3 N }
        \right)^{1/3}
    \right)
    \left( 1 + O \left(\sqrt{h^3 N}/t^3 \right) \right).
\end{multline*}
Consequently,
\begin{multline*}
    t
    = F_\beta^{-1}
       \left\{
           1 - \frac{\ce}{\ci}
           +
           \frac{\ce}{\ci}
           \sqrt{\frac{2}{\beta \pi}}
           \left(
               \frac{ h^3 N }{\gamma^{1/4} + \gamma^{-1/4}}
           \right)^{1/6}
    \right.
    \\
    \cdot
    \left . t^{-1}
           \exp \!
           \left(
               -\frac{\beta t^2}{8}
               \left(
                   \frac{\gamma^{1/4} + \gamma^{-1/4}}{ h^3 N }
               \right)^{\!1/3}
           \right) \!
           \left(
               1 + O\left(\sqrt{h^3 N}/t^3 \right)
           \right) \!
       \right\}
    \\ \!\!\!
    = t_{\sqrt{\gamma}}
       +
       \left[
           f_\beta \left( t_{\sqrt{\gamma}} \right)
       \right]^{-1}
       \cdot
       \frac{\ce}{\ci}
       \sqrt{\frac{2}{\beta \pi}}
       \left(
           \frac{ h^3 N }{\gamma^{1/4} + \gamma^{-1/4}}
       \right)^{1/6}
    \\
    \cdot t_{\sqrt{\gamma}}^{-1}
       \exp
       \left(
           -\frac{\beta t_{\sqrt{\gamma}}^2}{8}
           \left(
               \frac{\gamma^{1/4} + \gamma^{-1/4}}{ h^3 N }
           \right)^{1/3}
       \right)
       \left(
           1 + O\left(\sqrt{h^3 N} \right)
       \right) ,
\end{multline*}
where $t_{\sqrt{\gamma}} = F_\beta^{-1} \left( 1 - \frac{\ce}{\ci} \right)$.
Likewise, this expression only makes sense when
$\ce < (1 - F_\beta(0)) \cdot \ci$.  The last case we need to consider is
when $\ce = (1 - F_\beta(0)) \cdot \ci$.  In this case we have that $t$
solves
\begin{multline*}
    t^2
    =
    \left[
           f_\beta ( 0 )
    \right]^{-1}
    \cdot
    \frac{\ce}{\ci}
    \sqrt{\frac{2}{\beta \pi}}
    \left(
        \frac{ h^3 N }{\gamma^{1/4} + \gamma^{-1/4}}
    \right)^{1/6}
    \\
    \cdot   \exp
       \left(
           -\frac{\beta t^2}{8}
           \left(
               \frac{\gamma^{1/4} + \gamma^{-1/4}}{ h^3 N }
           \right)^{1/3}
       \right)
       \left(
           1 + O\left(\sqrt{h^3 N} \right)
       \right).
\end{multline*}
\end{IEEEproof}

\begin{IEEEproof}[Proof of Lemma~\ref{L:thresh-med-h}]
We now suppose instead that $h = h_0 N^{-1/3}$, for some constant
$h_0 > 0$.  In this case
\begin{multline*}
    \frac{ T - \mu_{N,n}(\sqrt{\gamma} + h)}
         { \sigma_{N,n}(\sqrt{\gamma} + h) }
    =
    \frac{\beta^{1/2} t}{2 \sqrt{ h_0 } }
    \left( \gamma^{1/4} + \gamma^{-1/4} \right)^{1/6}
    \\
    -
    \frac{\beta^{1/2} h_0^{3/2}}
         { \sqrt{ 2 \gamma } \left( \gamma^{1/4} + \gamma^{-1/4} \right) }
    +
    O( h_0 N^{-5/6 }).
\end{multline*}
If $\ce = \omega( \ci )$, then we must have $t \to -\infty$ so that the
left-hand side of~\eqref{E:T-eqn} converges to $\ci$.  Using the tail
behavior of $\Phi( \cdot )$, we have that
\[
    \frac{\beta^{1/2} t}{2 \sqrt{h_0}}
    \left(
        \gamma^{1/4} + \gamma^{-1/4}
    \right)^{1/6}
    \sim
    \Phi^{-1} \left( \frac{\ci}{\ce} \right)
    \sim
    -
    \sqrt{
        2 \log \frac{\ce}{\ci}
    }
\]
so that
\[
    t
    \sim
    -
    \sqrt{
        \frac{ 8 h_0 }{ \beta (\gamma^{1/4} + \gamma^{-1/4} )^{1/6} }
        \log \frac{\ce}{\ci}
    }.
\]
If, on the other hand, $\ce = o(\ci)$, then the right-hand side of~\eqref{E:T-eqn}
must converge to $\ce$, and
\[
    t
    \sim
    F_\beta^{-1} \left( 1 - \frac{\ce}{\ci} \right)
    \sim
    \left(
        \frac{3}{2 \beta}
        \log \frac{\ci}{\ce}
    \right)^{2/3}.
\]
\end{IEEEproof}

\section*{Acknowledgment}
The authors wish to thank Art Owen for many helpful discussions.

\bibliographystyle{IEEEtran}%
\bibliography{PerryWolfeRankEst09}%

\end{document}